# Does the performance of TDD hold across software companies and premises? A group of industrial experiments on TDD

Adrian Santos[1], Janne Järvinen[2], Jari Partanen[3], Markku Oivo[1], and Natalia Juristo[4]

[1] M3S (M-Group), ITEE University of Oulu, Finland,
`adrian.santos.parrilla/markku.oivo@oulu.fi`,
[2] F-Secure, Finland,
`janne.jarvinen@f-secure.com`
[3] Bittium, Finland,
`jari.partanen@bittium.com`
[4] Escuela Técnica Superior de Ingenieros Informáticos, Universidad Politécnica de Madrid, Spain,
`natalia@fi.upm.es`

**Abstract.** Test-Driven Development (TDD) has been claimed to increase external software quality. However, the extent to which TDD increases external quality has been seldom studied in industrial experiments. We conduct four industrial experiments in two different companies to evaluate the performance of TDD on external quality. We study whether the performance of TDD holds across premises within the same company and across companies. We identify participant-level characteristics impacting results. Iterative-Test Last (ITL), the reverse approach of TDD, outperforms TDD in three out of four premises. ITL outperforms TDD in both companies. The larger the experience with unit testing and testing tools, the larger the difference in performance between ITL and TDD (in favour of ITL). Technological environment (i.e., programming language and testing tool) seems not to impact results. Evaluating participant-level characteristics impacting results in industrial experiments may ease the understanding of the performance of TDD in realistic settings.

## 1 Introduction

TDD is an agile development approach that enforces the construction of software systems by means of short and iterative testing-coding cycles —contrary to traditional approaches, where coding is usually performed before testing, coding and testing are rarely interleaved, and testing is commonly performed after the whole system has been developed [1]. These short and iterative testing-coding cycles are, according to its proponents [1], the main reason behind TDD's superiority over traditional approaches (e.g., Waterfall) with regard to software quality. Even though the performance of TDD on various software quality attributes [2] has been studied before [3, 4, 5, 6, 7, 8, 9], external quality seems to be the most investigated so far. External quality is usually considered in the TDD literature as the proportion of test cases that successfully pass from a battery of tests specifically built for testing the software system under development.[1]

Several industrial case studies and surveys support the superiority of TDD over traditional approaches with regard to external software quality [3, 4, 5, 6]. However, the extent to which TDD outperforms control approaches with regard to external quality varies largely from study to study [10, 11]. This may be due to the technological environments on which studies are run or due the characteristics of the subjects participating in the studies (e.g., professional experience, skills, background, etc.). Unfortunately, despite the alleged benefits of industrial experiments (e.g., making causality claims on technology performance in realistic settings [12, 13], increasing internal validity compared to industrial case studies or surveys [14], etc.), only two of the studies conducted so far on TDD —evaluating external quality— are industrial experiments (i.e., [15, 16]). Unfortunately, in none of them it is possible isolating the effects of TDD on external quality. This led Munir et al. to claim in one of the latest secondary studies conducted on TDD [3]: *"strong indications are obtained that external quality is positively influenced, which has to be further substantiated by industry experiments..."*.

Along this article we aim to answer the following **research questions** with regard to the performance of TDD on external quality:

– Does TDD outperform control approaches in industrial experiments as in case studies and surveys?
– Does the performance of TDD hold across premises within the same company and across companies?

To answer these questions we conduct a group of four industrial experiments evaluating the effects of TDD and ITL on external quality. We run three experiments at F-Secure —a multinational security and digital privacy company [17]— and one at Bittium —a multinational telecommunications company. We first analyze all the experiments individually, and then, we combine their results by means of meta-analysis [18]. Finally, we assess the extent to which results hold across premises within the same company and across companies and identify participant-level characteristics that may be behind the variability of results observed. Throughout this research, we made several **findings**:

---

[1] For simplicity's sake, along the rest of the article we refer to external quality and quality interchangeably.

---

**Key findings**

- According to our results, ITL outperforms TDD for novices on TDD. Results hold across the two companies that we have studied.
- The extent to which ITL outperforms TDD looks dependent upon participant-level characteristics. In particular, the larger the experience with unit testing and testing tools, the larger the difference in performance between ITL and TDD (in favour of ITL).
- ITL outperforms TDD in our group of industrial experiments. This is contrary to what has been previously claimed in case studies and surveys. This difference of results may have emerged due to the lack of previous familiarity of our participants with the TDD process.

---

The main **contributions** of this paper are a *comparison of the results achieved in four industrial experiments on TDD* and *the first assessment of participant-level characteristics impacting the performance of TDD across software industries*. As a secondary contribution we offer a compilation of the primary studies that evaluate the effects of TDD on external quality in industry and a comparison of F-Secure and Bittium's results with those.

Along this study we argue that despite the long years of research on TDD, almost none of the available studies has evaluated the effects of TDD on quality in industrial experiments. Industrial experiments not only allow to assess the effects of TDD on quality in realistic environments, but also, the effect of practitioners' characteristics on TDD's performance. In view of this, we suggest:

---

**Actionable results**

- The impact of *participant-level characteristics* (e.g., experience with programming, unit testing, etc.) should be studied to learn about the practitioners' characteristics that impact TDD's performance.
- As industrial experiments' sample sizes tend to be small, *replications* shall be conducted and analyzed jointly to detect participant-level characteristics impacting results.
- Participants' previous familiarity with more traditional development approaches (e.g., ITL) may distort the evaluation of TDD's performance. In view of this, we suggest to assess the performance of TDD in further occasions in industrial *between-subjects experiments* —being the subjects in each group either experts in TDD, or experts in ITL, respectively.

---

**Paper organization**. In Section 2 we portray the related work of this study. In Section 3 we outline the characteristics of our group of experiments. In Section 4 we conduct the analysis of our group of experiments. Then, in Section 5 we discuss the results of our group of experiments and put them in perspective.

Finally, in Section 6 we outline the threats to validity of our study, and then in Section 7 the conclusions.

## 2 Related Work

To gather a list of primary studies evaluating the effects of TDD on external quality in industrial settings, we go over the secondary studies conducted so far on TDD [3, 4, 5, 6, 7, 8, 9]. Table 1 shows the list of the primary studies that we identified, their research methods (i.e., case studies, surveys or controlled experiments following Wohlin et al. definitions [14]) and their results (i.e., the difference in performance between TDD and the control approach).

**Table 1.** TDD effects on quality in industrial studies.

| Method | Reference | Result |
|---|---|---|
| Case study | [19][20][21][22][23][24][25][26][27][10][28][11][29] | + |
| Survey | [30][31] | + |
| Experiment | [15] | + |
| | [16] | ? |

As it can be seen in Table 1, all studies —but one experiment [16]— report positive results (i.e., TDD is superior to the control approach).

Even though all the *case studies* report that TDD outperforms control approaches with regard to quality, wildly heterogeneous improvements with TDD over control approaches are claimed [3, 4, 5, 6, 7, 8, 9]: ranging from improvements as low as 18% [10], to improvements as high as 50% [11, 25]. Such heterogeneity of results may have emerged due the different technological environments where case studies were run or the characteristics of the participants involved: from environments where JUnit and Java were used to develop systems from scratch by accountants, lecturers or expert programmers [10] to environments where C++ and CUnitTest were used to increase the functionality of legacy systems by groups of intermediate and novice developers [21].

With regard to the evidence obtained in *surveys*, one found that TDD outperformed industry averages on quality [31] and another found that TDD helps to achieve greater quality than control approaches [30]. Again, participants' characteristics varied largely across studies: from largely experienced practitioners [31] to developers with almost no previous experience in programming [30].

Despite their advantages for obtaining preliminary evidence [14], case studies and surveys are usually included within the lowest positions in hierarchies of evidence [32, 33] due to their inability to prove causality. This is so because in such empirical studies other elements rather than the technologies themselves may be the cause of results (e.g., external factors in case studies or personal opinions in surveys). This is where experiments have their natural fit [14].

From the two *experiments* conducted so far in industry [15, 16], the first experiment [16] reports inconclusive results (as all subjects are able to achieved the maximum quality regardless of the development approach being applied). The second experiment [15] evaluated the effects of TDD and pair-programming together and thus, the effects of TDD cannot be isolated from those of pair-programming (i.e., TDD and pair-programming effects are confounded).

As a summary of the evidence collected so far on TDD with regard to external quality in industrial settings, most studies are either surveys or case studies. In those, a large heterogeneity of results materialized: either due to differences across technology environments or participants' characteristics. Besides, in the pair of experiments available, certain shortcomings did not allow to assess the extent to which TDD affects quality. In turn, the question of how does TDD influence quality in industrial experiments is still unanswered, since the only type of empirical study able to prove causality are experiments [14].

We previously ran a series of *identical experiments* at F-Secure [17] and got opposite results to those obtained in industrial case studies and surveys. In particular, according to the results obtained at F-Secure, ITL outperformed TDD. However, F-Secure's experiments are just the first-step towards proving causality. In particular, F-Secure's experiments' results may be artifactual (i.e., caused by the technological environment of the experiment) or have occurred just by chance [34]. Thus, in this study we run a replication of F-Secure's experiment at a different company (i.e., Bittium) changing the technological environment to validate our previous results. In addition, we meta-analyze their results together with the aim of increasing the reliability of the joint findings and study the differences of results across companies.

## 3 Group of Experiments

We conducted a total of four experiments to evaluate the effects of TDD on quality. Three *exact replications* were run at F-Secure (each one at a different location: Helsinki, Kuala-Lumpur and Oulu). Differences across F-Secure's experiments' results and the participant-level characteristics that may have led to such differences were not investigated before. We conducted a *close replication* at Bittium. We introduced as few changes as required by Bittium's managers (i.e., changes in the programming language and the testing tool) with the aim of minimizing the risk of confounding effects across experiments. Thus, Bittium's experiment is a *close replication* of F-Secure's. This should increase the reliability of the joint findings, ease the comparison of results across companies, and at the same time, facilitate the elicitation of practitioners' characteristics impacting TDD's performance across software industries.

### 3.1 Dependent and independent variables

The independent variable across all the experiments is **development approach**, with TDD and ITL as treatments. ITL was defined as the reverse-order approach of TDD following Erdogmus et al. [35].

The dependent variable across all the experiments is **external quality**. As commonly done in the TDD literature, we measure external quality as the *percentage* of test cases that successfully pass from a battery of tests that we specifically built for testing participants' solutions. Specifically, we measure external quality as:

$$QLTY = \frac{\#Test\ Cases(Pass)}{\#Test\ Cases(All)} * 100$$

### 3.2 Experimental settings

Seminars on TDD were conducted at each site. An experiment was embedded within each seminar. Table 2 summarizes the settings of the experiments conducted at F-Secure and Bittium (changes across companies' settings in *italics*).

**Table 2.** Experiments' settings: F-Secure and Bittium.

| Aspect | Values |
|---|---|
| **Factors** | Development Approach |
| **Treatments** | TDD vs ITL |
| **Response variables** | QLTY |
| **Design** | AB Repeated-measures |
| **Training** | TDD course |
| **Training duration** | 3 days/6 hours |
| **Experiment Duration** | 2.25 hours |
| **Programming Language** | F-Secure: *Java*; Bittium: *C++* |
| **Unit Testing Tool** | F-Secure: *JUnit*; Bittium: *GTest* |

The trade-off assessment of the experimental design, the specification of the instruments and the experimental tasks, and an in-depth discussion of the threats to validity of F-Secure's experiments can be found elsewhere [17].

### 3.3 Subjects

Subjects were handed a survey some days before the experiment. The survey contained a series of ordinal-scale (i.e., inexperienced, novice, intermediate and expert) self-assessment questions with regard to their experience with programming, unit testing and the programming language and testing tool used during the experiment.[2] Table 3 shows the mean —and standard deviations— of the participants' experiences with programming, the programming language, unit testing and the testing tool used within the experiment (1-4, for inexperienced, novice, intermediate and experts, respectively).[3]

---

[2] The survey and its results were published elsewhere [36].

**Table 3.** Mean and standard deviation of experiences across experiments.

| Experiment | N | Programming | Prog. Language | Unit Testing | Testing Tool |
|---|---|---|---|---|---|
| F-Secure H | 6 | 3.67 (0.52) | 2.33 (1.21) | 2.17 (0.98) | 2.17 (1.17) |
| F-Secure K | 11 | 2.91 (0.7) | 1.82 (0.87) | 1.64 (0.5) | 1.27 (0.47) |
| F-Secure O | 7 | 3.29 (0.76) | 2.71 (1.11) | 2.71 (0.76) | 2 (0.82) |
| Bittium | 9 | 3 (0.87) | 2.89 (0.93) | 1.67 (0.87) | 1 (0) |

To ease the interpretation of the data presented in Table 3, we provide in Figure 1 a profile-plot showing the mean of the experiences of the participants in each experiment.

**Fig. 1.** Profile-plot for experiences across experiments.

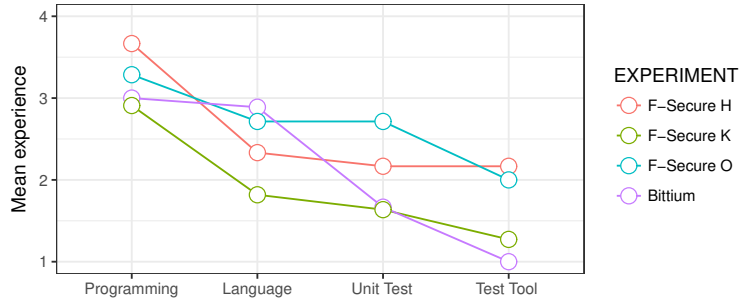

As it can be seen in Figure 1, F-Secure H's participants are the most experienced with programming. Besides, F-Secure O and Bittium's participants seem the most experienced with the programming language used during the experiment. Finally, subjects across all experiments have lower experiences with unit testing and testing tools than with programming or the programming language used during the experiment. In general, those with the higher experience in unit testing and testing tools tend to be the most senior professionals. Overall, our group of experiments is formed by *an heterogeneous population of TDD novices with relatively low experience in unit testing and testing tools.*

---

[3] For simplicity's sake, we consider the variables measured along the survey as continuous. This approach is commonly followed in other disciplines [37].

### 3.4 Analysis Approach

First, we provide the *descriptive statistics* (i.e., number of data-points, mean, median and standard deviation) of ITL and TDD across the experiments. Then, we provide a profile plot for easing the understanding of the data.

Afterwards, we analyze each experiment *individually* with an identical statistical test: a repeated measures analysis of variance (RM ANOVA) [38]. The RM ANOVA assumes that the residuals are normally distributed [38]. We check the normality assumption by means of the Shapiro-Wilk test [38].

Then, with the aim of providing a *joint result*, we combine the results of the RM ANOVAs jointly by means of a random-effects meta-analysis following the steps outlined by Burke et al. [39]. We also perform a *sub-group meta-analysis* [18] to assess the extent to which results hold across companies (i.e., F-Secure vs. Bittium).

Finally, with the aim of identifying participant-level characteristics influencing results, we perform a series of "post hoc" analyses —one per experience variable (i.e., programming experience, programming language experience, unit testing experience and testing tool experience). We follow Fisher et al.'s guidelines [40] to identify participant-level characteristics impacting results.

## 4 Analysis

### 4.1 Descriptive Statistics

Table 4 shows the descriptive statistics (i.e., sample size, mean, standard deviations and medians) for ITL and TDD across all experiments. To ease the interpretation of the data presented in Table 4, Figure 2 shows the profile-plot of the means for ITL and TDD across experiments.

**Table 4.** Descriptive statistics for quality: ITL vs TDD across experiments.

| Experiment | Treatment | N | Mean | SD | Median |
|------------|-----------|---|------|-----|--------|
| F-Secure H | ITL | 6 | 30.71 | 36.58 | 24.16 |
|            | TDD | 5 | 18.48 | 7.30 | 16.67 |
| F-Secure K | ITL | 11 | 22.17 | 20.44 | 17.98 |
|            | TDD | 11 | 13.98 | 10.21 | 13.64 |
| F-Secure O | ITL | 7 | 16.05 | 20.81 | 7.87 |
|            | TDD | 7 | 16.99 | 15.08 | 19.70 |
| Bittium    | ITL | 9 | 15.45 | 18.52 | 5.75 |
|            | TDD | 9 | 2.47 | 0.26 | 2.38 |

As it can be seen in Table 4, TDD's mean scores go from as high as $M = 18.48$ in F-Secure H to as low as $M = 2.47$ in Bittium. Bittium's participants achieved the lower ITL's mean scores ($M = 15.45$) while F-Secure H obtained the largest ($M = 30.71$). Moreover, while F-Secure H, F-Secure K and Bittium's participants showed lower mean performance with TDD than with ITL (i.e., negative slope in Figure 2), F-Secure O participants show an almost similar performance with TDD and ITL (even though TDD slightly outperforms ITL).

**Fig. 2.** Profile-plot for ITL and TDD across experiments.

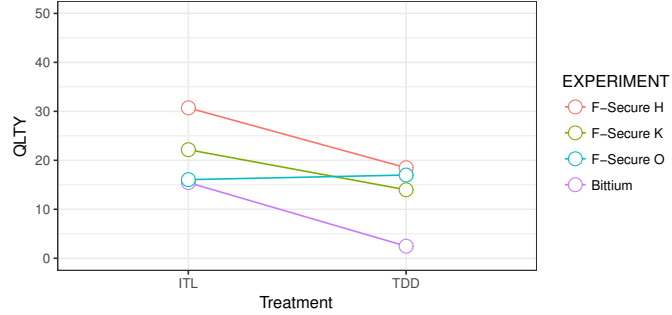

### 4.2 Individual Analyses

Table 5 shows the results of analyzing each experiment with a RM ANOVA.[4]

**Table 5.** Results across experiments.

| Experiment | N | Estimate | SE | $t$-value | $p$-value | Significance |
|------------|---|----------|------|-----------|-----------|--------------|
| F-Secure H | 5 | -14.12 | 14.17 | -0.99 | 0.345 | ✗ |
| F-Secure K | 11 | -8.18 | 5.98 | -1.37 | 0.186 | ✗ |
| F-Secure O | 7 | 0.93 | 5.68 | 0.16 | 0.871 | ✗ |
| Bittium | 9 | -12.98 | 6.20 | -2.09 | 0.053 | ✗ |

As it can be seen in Table 5, ITL outperforms TDD in three out of four experiments (as treatment estimates are negative in F-Secure H, F-Secure K and Bittium). In addition, the difference in performance between them is not statistically significant in any experiment (even though Bittium's is almost statistically significant).

### 4.3 Joint Result and Sub-Group Meta-Analysis

After analyzing each experiment individually, we aggregate their results by means of a random effects meta-analysis [18]. Figure 3 shows the results of the meta-analysis performed (and the results of the sub-group meta-analysis performed for each company).

---

[4] The normality assumption is met in all experiments according to the Shapiro-Wilk test [38].

**Fig. 3.** Forest-plot: TDD vs ITL.

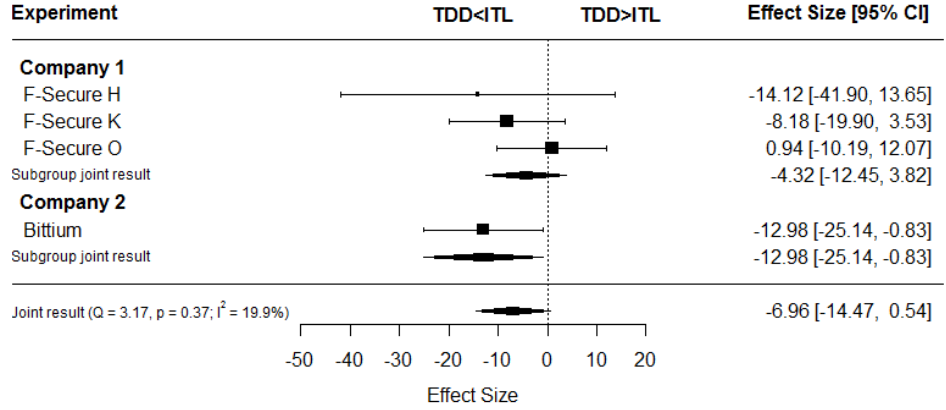

According to the joint result (see the black diamond at the bottom of Figure 3), *ITL outperforms TDD to a small* ($M = -6.96$; $95\%CI = (-14.47, 0.54)$) and *non-statistically significant extent* (as the 95% CI crosses 0). Besides, Bittium's results seem within the realm of those of F-Secure (as Bittium's effect size is even smaller in *magnitude* than that of F-Secure H). In addition, notice how F-Secure's joint effect size ($M = -4.32$; $95\%CI = (-12.45, 3.82)$) overlaps with that of Bittium ($M = -12.98$; $95\%CI = (-12.45, 3.82)$). In view of this, *results hold across companies: TDD does not increase quality (compared to ITL) in none of the companies.* In turn, as results hold across companies, we conclude, *technological environment* (e.g., Java and JUnit at F-Secure and C++ and GTest at Bittium) *seems not to impact results.*

Finally, even though results hold across companies, *heterogeneity emerged* when aggregating results together ($I^2 = 19.9\%$). As heterogeneous results emerged even among the identical experiments conducted at F-Secure (see that F-Secure O's results are observably different that those of F-Secure H and K), we hypothesize, *participant-level characteristics* (i.e., experience with programming, programming languages, unit testing or testing tools) *may be behind the heterogeneity of results detected.*

### 4.4 Post-hoc Analysis: Developers' Characteristics

We performed four different RM ANOVAs (one per experience variable) to assess the effects of participant-level experiences on results. Table 6 shows the estimates and standard errors provided by the RM ANOVAs fitted to assess the effects of the experience variables on results.

Table 6 can be read as the impact of one unit increase in experience on the performance achieved with TDD *beyond the performance achieved with ITL*. For example, per each unit increase in experience with the testing tool used (see the row "Testing Tool" in Table 6), the performance with TDD decreases in $M = -6.80; SE = 5.66$ units. Thus, the larger the experience with the testing tool (ranging between 1 to 4), the lower the performance with TDD in comparison with the performance achieved with ITL. Notice how in comparison with the joint result of our group of experiments (i.e., $M = -6.96$ according to the results achieved in the previous section), the decrease in quality per unit increase in experience with the testing tool seems considerable (i.e., almost a 1:1 ratio). Thus, experience with the testing tool seems to be a relevant *moderator* of the effects of TDD. With the aim of easing the understanding of results, Figure 4 shows the regression lines corresponding to the estimates of Table 6.[5]

**Table 6.** Participant-level characteristics impact on TDD's performance.

| Factor | Estimate | SE | $t$-value | $p$-value | Significance |
|--------|----------|-----|-----------|-----------|--------------|
| Programming | 0.74 | 5.32 | 0.13 | 0.89 | ✗ |
| Language | 0.78 | 3.94 | 0.19 | 0.84 | ✗ |
| Unit Testing | -3.54 | 5.09 | -0.69 | 0.49 | ✗ |
| Testing Tool | -6.80 | 5.66 | -1.19 | 0.24 | ✗ |

**Fig. 4.** Participant-level characteristics impact on TDD's performance.

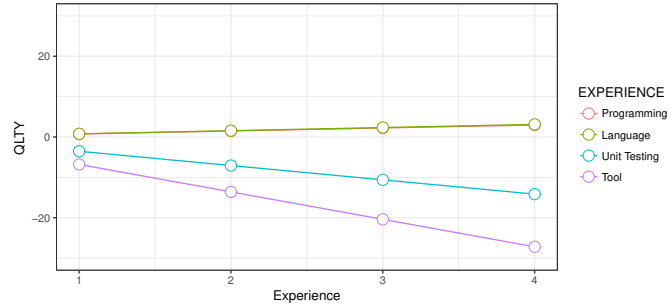

As it can be seen in Figure 4, the experience with programming or programming languages seems not to impact results (as lines seem flat along all the experience levels). However, *the larger the experience with unit testing or the testing tool, the lower the performance with TDD* in comparison with the performance achieved with ITL (as the lines have a downward slope).

## 5 Discussion

ITL outperformed TDD in three out of four experiments (i.e., F-Secure H, F-Secure K and Bittium). Besides, after aggregating their results together, still

---

[5] The regression lines for "Programming" and "Language" partially overlap and thus, only that of "Language" is visible.

ITL slightly outperformed TDD. The results of our group of experiments cannot be compared with those of any other industrial experiment (as the couple of experiments published so far do not allow to isolate the effects of TDD on quality). However, our results can be compared with those of industrial case studies and surveys [3, 4, 5, 6, 7, 8, 9]. Long story short, our results are opposite to those of case studies and surveys: while we observed a slight decrease in the performance achieved with TDD over ITL, surveys and case studies reported large improvements with TDD over control approaches. These opposite results may have emerged as our participants had no previous experience on TDD, and in turn, this may have lowered their performance with TDD in comparison with more "traditional" approaches such as ITL.

Our results also suggest that the larger the experience with unit testing and testing tools (and thus, the potentially larger the experience with more traditional approaches including testing such as ITL), the larger the drop in quality when applying TDD (the recently learned development approach). In view of this, we suggest that for making an objective evaluation of TDD's performance in industrial experiments, it may be necessary to run between-subjects experiments with two similarly experienced treatment groups (e.g., experts in TDD vs. experts in ITL). Otherwise, the development approach in which the participants are more experienced may end up being favoured.

Finally, the different technological environments used in our group of experiments (i.e., Java and JUnit for F-Secure, and C++ and GTest for Bittium), seem not to impact results (as results hold across companies). In view of this, and also in view of the large heterogeneity of results observed across case studies and surveys, we hypothesize, participant-level characteristics may be behind the large heterogeneity of results observed in literature.

## 6 Threats to Validity

**Conclusion validity** concerns the statistical analysis of results [14]. We used commonly applied statistical techniques to analyze the data (i.e., RM ANOVA [38] and random-effects meta-analysis [18]). In addition, the required data assumptions (i.e., normality assumption [38]) were assessed before interpreting the results of the RM ANOVA. In view of this, we do not expect any threat to conclusion validity to impact our results.

**Internal validity** is the extent to which the detected effects are caused by the treatments and not by other variables beyond researcher's control [14]. There is a potential maturation threat: seminars were offered as a three-day training course on TDD and contained multiple exercises and experimental laboratories. Thus, factors such as tiredness or inattention might have materialized. In order to minimize this threat we ensured that subjects were given enough breaks. Treatment leakage may have influenced results: subjects might have increased their performance with TDD because they learned something (i.e., slicing) in the previous session while applying ITL. However, treatment leakage seems not to have materialized: ITL outperformed TDD in three out of four experiments.

**Construct validity** refers to the correctness in the mapping between the theoretical constructs under investigation and the operationalizations of the variables in the study. As usual in SE experiments, the study suffers from the mono-operation bias (as only test cases were used to measure external quality). Conformance to the development approaches is one of the notorious threats to validity in most SE experiments. This threat was minimized by encouraging subjects to adhere as closely as possible to the development approaches taught during the seminar. We complemented such encouragement with visual supervision.

**External validity** relates to the possibility of generalizing results beyond the objects and subjects involved in the study [14]. As usual in SE experiments, it was not possible to obtain random samples from the population under study: convenience sampling was used in all experiments. Experiments were conducted with toy-tasks. This may limit the generalizability of the findings. However, we expect results to be representative for professionals who are starting to learn the TDD process coding toy-tasks.

## 7 Conclusion

TDD has been claimed to increase external quality over traditional approaches in industrial case studies and surveys [3, 4, 5, 6, 7, 8, 9]. However, the extent to which TDD performs in industrial experiments has been seldom studied. Industrial experiments allow to assess not just the performance of TDD in realistic settings, but also the effects of participant-level characteristics on TDD's performance.

We conducted a group of four industrial experiments with TDD novices to assess the extent to which TDD performs in industrial settings. ITL slightly outperformed TDD in three out of four experiments. When aggregating their results together by means of meta-analysis, ITL still slightly outperformed TDD. Our results are opposite to those found in industrial case studies and surveys. These different results may have emerged due to the lack of previous familiarity of our experiments' participants with the TDD process.

Finally, results held across the two companies that we studied. In view of this, companies' technological environments seemed not to impact results. However, the extent to which ITL outperformed TDD looked dependent upon participants' characteristics (as heterogeneity materialized even across identical experiments). According to our results, the larger the experience with unit testing and testing tools (and thus, the potentially larger experience with more traditional approaches), the larger the drop in performance with TDD over the performance achieved with ITL. In view of this, we suggest, the development approach in which subjects are more experienced may end up being favoured in controlled experiments. With the aim of tackling this shortcoming, we suggest to run between-subjects experiments evaluating the performance of TDD and control approaches with similarly experienced subjects across groups.

Once again in SE [41], a large number of elements seem to impact the performance of a technology (in this case TDD) in realistic settings. In this research we want to make a call to run experiments in industry and to assess not just the effects of TDD, but also, disentangle the characteristics of the participants that make TDD more or less desirable. Unfortunately, as SE industrial experiments tend to be small, individual experiments may be under-powered for detecting participant-level characteristics impacting results [42]. In view of this, multiple replications may be required to be conducted and analyzed jointly (as access to the raw-data is a requirement to elicit participant-level characteristics [43]). Are we still going to think that one size fits all? Or instead, are we going to learn about the characteristics that make practitioners more prone to benefiting from the application of TDD? After all, we are still at the beginning of understanding how TDD works in industrial settings [44].

## Acknowledgments

This research was developed with the support of the Spanish Ministry of Science and Innovation project TIN2014-60490-P.